\documentclass{article}
\usepackage{amsmath}
\usepackage{graphicx,graphics,rotating,multirow,lscape,setspace}
\usepackage{authblk}
\usepackage[left=2cm,right=2cm,vmargin=2cm]{geometry}
\title{Multiple imputation of covariates by fully conditional specification: accommodating the substantive model}

\author[1]{Jonathan W. Bartlett}
\author[2]{Shaun R. Seaman}
\author[2]{Ian R. White}
\author[1]{James R. Carpenter}
\affil[1]{Department of Medical Statistics, London School of Hygiene \& Tropical Medicine, UK}
\affil[2]{MRC Biostatistics Unit, Cambridge, UK\\

and for the Alzheimer's Disease Neuroimaging Initiative investigators*}

\newcommand{\iid}{\stackrel{\mathrm{iid}}{\sim}}
\newcommand{\etal}{\textit{et al }}
\newcommand{\cind}{\mbox{$\perp\hspace{-0.5em}\perp$}}

\newcommand{\expit}{\mbox{expit}}
\newcommand{\Var}{\mbox{Var}}

\begin{document}

\maketitle

\begin{abstract}
Missing covariate data commonly occur in epidemiological and clinical research, and are often dealt with using multiple imputation (MI). Imputation of partially observed covariates is complicated if the substantive model is non-linear (e.g. Cox proportional hazards model), or contains non-linear (e.g. squared) or interaction terms, and standard software implementations of MI may impute covariates from models that are incompatible with such substantive models. We show how imputation by fully conditional specification, a popular approach for performing MI, can be modified so that covariates are imputed from models which are compatible with the substantive model. We investigate through simulation the performance of this proposal, and compare it to existing approaches. Simulation results suggest our proposal gives consistent estimates for a range of common substantive models, including models which contain non-linear covariate effects or interactions, provided data are missing at random and the assumed imputation models are correctly specified and mutually compatible.
\end{abstract}


*Data used in preparation of this article were obtained from the Alzheimer's Disease Neuroimaging Initiative
(ADNI) database (adni.loni.ucla.edu). As such, the investigators within the ADNI contributed to the design
and implementation of ADNI and/or provided data but did not participate in analysis or writing of this report.
A complete listing of ADNI investigators can be found at:\\
http://adni.loni.ucla.edu/wp-content/uploads/how\_to\_apply/ADNI\_Acknowledgement\_List.pdf

\section{Introduction}
\label{introduction}
Missing data is a pervasive problem in both experimental and observational medical research, causing a loss of information and potentially biasing inferences. In this article we focus on settings in which interest lies in fitting a substantive model relating an outcome to a number of covariates, one or more of which contain missing values. The method of multiple imputation (MI) has become an extremely popular approach for accommodating missing data in statistical analyses, both generally \cite{Kenward/Carpenter:2007}, and in the specific context of partially observed covariates \cite{White2010a}. MI involves `filling in' each missing value with draws from an appropriate distribution, leading to a number $M$ of completed datasets. The substantive model can then be fitted to each of the $M$ completed datasets, and the results combined across the $M$ datasets using Rubin's rules \cite{Rubin:1987}, which account for the uncertainty due to the fact data have been imputed. MI is most often applied under the missing at random (MAR) assumption, which stipulates that the probability that data are missing are independent of the missing values, conditional on the observed data, although MI can also be used when data are missing not at random \cite{Rubin:1987}.

Parametric MI as originally proposed is based on a joint imputation model for the partially observed variables (conditional on any fully observed variables), which we refer to as `joint model MI'. A popular alternative to joint model MI is the fully conditional specification (FCS) approach (\cite{White2011}, \cite{Buuren:2007}). FCS MI involves specifying a series of univariate models for the conditional distribution of each partially observed variable given the other variables. This permits a great deal of flexibility, since an appropriate regression model can be selected for each variable (e.g. linear regression for continuous variables, logistic regression for binary variables). Consequently, FCS MI is particularly appealing in settings in which a number of variables have missing data, some of which are continuous and some of which are discrete.

One of the strengths of MI is that it divides the process of dealing with missingness (the imputation stage) from the analysis of the completed data (the analysis stage). As has been previously discussed in detail, this division presents both opportunities and threats (\cite{Meng:1994,Schafer:1997,Collins2001,Schafer2003}). For example, we may be able to include so called auxiliary variables in the imputation model which are not used in the analysis stage. This offers the potential for increased efficiency and may also improve the plausibility of the MAR assumption holding, by conditioning on auxiliary variables which are predictive of missingness. Sometimes the imputer and analyst will be different people. If the imputer has additional knowledge which enables them to impose (correct) additional assumptions in the imputation model, the analyst will gain efficiency.

The division may however sometimes lead to problems. In the context of imputing partially observed covariates, imputations might be generated from a model which is \textit{incompatible} with the substantive model, which may lead to (asymptotically) biased estimates of parameters in the latter. Two conditional models are said to be incompatible if there exists no joint model for which the conditionals (for the relevant variables) equal these conditional models. This is a particular issue when the substantive model contains non-linear covariate effects or interactions, with which default imputation model choices may be incompatible. For example, suppose the substantive model is the linear regression of $Y$ on a continuous covariate $X$ and $X^{2}$, and we wish to impute missing values in $X$. The default choice for the imputation model for $X|Y$ in software for MI is a normal linear model, with conditional mean equal to a linear function of $Y$, which is incompatible with the quadratic substantive model. Following imputation of $X$, $X^{2}$ is then passively imputed by squaring the imputed $X$ values. In this case, estimates of the parameters of the substantive model from multiply imputed datasets will be biased (unless the quadratic coefficient is in truth zero), because within the subset of data where $X$ has been imputed the association between $X$ and $Y$ is linear as a consequence assuming linearity in the imputation model \cite{Seaman2012}.

Incompatibility between the imputation and substantive models can be avoided by specifying a joint model for outcome and covariates for which the conditional distribution of outcome given covariates matches the substantive model and then using the imputation model implied by this joint model. However, specification of a joint model is challenging when there are a number of partially observed covariates, particularly when some are continuous and some are discrete. In this setting the FCS method is an attractive option. However, the default univariate imputation models used in FCS may be incompatible with the substantive model. In this paper we therefore propose a modification of the popular FCS approach to MI which ensures that each of the univariate imputation models is compatible with the assumed substantive model.

We begin in Section \ref{adniIntro} by introducing a motivating example from the Alzheimer's Disease Neuroimaging Initiative (ADNI). In Section \ref{compatibility} we formally define compatibility between an imputation model for partially observed covariates and a substantive model, explain when incompatibility implies imputation model mis-specification, and give examples of when this occurs. We then outline how imputation models can be specified which are compatible with a given substantive model within the joint modelling approach to MI, which motivates our modification of the FCS MI approach. In Section \ref{fcs} we briefly review the standard FCS framework for MI in the setting of partially observed covariates. In Section \ref{accommodatingTheOutcomeModel} we describe our modification of the FCS MI approach which ensures that each univariate imputation model is compatible with the substantive model. We give details for how this can be done when the model of interest is i) normal linear regression, ii) a model for a discrete outcome (e.g. logistic and Poisson regression), or iii) a proportional hazards model. We report the results of a simulation study to investigate the performance of our proposed approach in Section \ref{simulationSection}. In Section \ref{adni} we apply our proposed approach to the motivating example. In Section \ref{exploratoryAnalyses} we discuss how our proposed approach can be used when, as is often the case, interest lies in fitting a number of different substantive models to the data. We conclude in Section \ref{discussion} with a discussion.

\section{Motivating example}
\label{adniIntro}
The Alzheimer's Disease Neuroimaging Initiative (ADNI) was launched in 2003 by the National Institute on Aging (NIA), the National Institute of Biomedical Imaging and Bioengineering (NIBIB), the Food and Drug Administration (FDA), private pharmaceutical companies, and nonprofit organizations, as a 5-year public-private partnership. The aims of ADNI included assessing the ability of imaging and other biomarkers to measure the progression of mild cognitive impairment and early Alzheimer's Disease (AD). The study aimed to recruit approximately 200 cognitively normal older individuals (controls), 400 with mild cognitive impairment (MCI), and 200 with early AD. Participants underwent clinical and cognitive assessment and MRI brain scans at baseline and at specified intervals (every 6 or 12 months, depending on subject group) up to 3 years. Further details regarding ADNI are given in the Acknowledgements.

Recently Jack Jr \etal used data from ADNI to investigate baseline predictors of time to conversion to AD in those subjects with MCI at baseline (\cite{Jack2010}). In particular, using Cox proportional hazards models they found evidence of a non-linear association between amyloid $\beta$ 1-42 peptides (A$\beta_{1-42}$) measured from cerebrospinal fluid (CSF) at baseline and hazard of conversion. They also found evidence that lower baseline hippocampal volume was predictive of increased hazard, after adjusting for total intracranial volume (a measure of head size). Participants were invited to have CSF measured at baseline by lumbar puncture, but were not required to do so to participate in the study. Consequently, only around 50\% of subjects had CSF A$\beta_{1-42}$ measured. The analysis of Jack Jr was restricted to the subset of $n=218$ MCI subjects for whom CSF A$\beta_{1-42}$ was measured. It may be reasonable to assume that the propensity to agree to lumbar puncture (and thus have CSF A$\beta_{1-42}$ measured) is independent of time to conversion to AD, conditional on CSF A$\beta_{1-42}$, and so such a complete case analysis might reasonably be assumed to be unbiased. However, it is inefficient, since it only uses data on 50\% of MCI subjects. Jack Jr \etal also found evidence that presence of the APOE4 gene, previously shown to be associated with development of AD in a number of studies, was associated with increased hazard for conversion to AD.

MCI is a heterogeneous classification, with only a certain proportion of subjects eventually going on to develop AD. For each subject their family history was collected at baseline, in particular in relation to whether their mother or father suffered from dementia or AD. Given that there is a genetic component to the disease, we were interested to investigate whether the presence of family history of AD was associated with increased hazard of conversion to AD, by including covariates indicating whether the subject's mother and father had had AD (Table \ref{adniBaseline}). Unfortunately, although family history of dementia was well recorded, family history of AD specifically suffered from missingness (see Table \ref{adniBaseline}).

We aimed to estimate the parameters of a Cox proportional hazards model for hazard of conversion to AD using the available data from the $n=382$ ADNI subjects who had MCI at baseline and who had at least one follow-up visit. Of these subjects, 167 were observed to convert to AD during follow-up. Our substantive model contained as covariates the variables listed in Table \ref{adniBaseline}, plus the square of A$\beta_{1-42}$ to allow for the non-linear association previously identified by \cite{Jack2010}. In addition to CSF A$\beta_{1-42}$, we include CSF Tau and P-tau as covariates, which are also thought to reflect pathology, and thus might be associated with hazard of conversion to AD. Tau and p-tau were log transformed to reduce skewness in their distribution. The FCS approach to MI is attractive here, since seven covariates are partially observed, with some continuous and some binary. However, we should be careful to ensure that the imputation models we use are compatible with the substantive model, which includes a quadratic effect of one of the partially observed covariates. If we impute from a model which does not allow for a potential non-linear association between (log) hazard and CSF A$\beta_{1-42}$, we would expect to obtain inconsistent parameter estimates, particularly of the coefficients relating to CSF A$\beta_{1-42}$.

\begin{table}
\caption{\label{adniBaseline}Baseline characteristics of $n=382$ ADNI subjects with MCI at baseline}
\centering
\fbox{%
\begin{tabular}{lll}
Variable & Mean (SD) or no. (\% of observed) & No. missing values (\%) \\
\hline
A$\beta_{1-42}$ (ng/mL) & 16.4 (5.5) & 190 (49.7 \%) \\
log(Tau) (log pg/mL) & 4.50 (0.49) & 193 (50.5 \%) \\
log(P-tau) (log pg/mL) & 3.44 (0.50) & 189 (49.5 \%) \\
Mother had AD & 77 (25.3 \%) & 77 (20.2 \%) \\
Father had AD & 26 (9.0 \%) & 93 (24.3 \%) \\
Intracranial volume (cm$^3$) & 1474 (150) & 43 (11.3 \%) \\
Hippocampal volume (cm$^3$) & 6.47 (1.04) & 43 (11.3 \%) \\
APOE4 positive & 207 (54.2 \%) & 0 (0 \%) \\
\end{tabular}}
\end{table}

\section{Multiple imputation of partially observed covariates}
\label{compatibility}
\subsection{Setup}
We consider the setting in which interest lies in fitting a model to a fully observed outcome $Y$ with $p$ partially observed covariates $X=(X_{1},..,X_{p})$ and $q$ fully observed covariates $Z=(Z_{1},..,Z_{q})$. Let $X^{obs}$ and $X^{mis}$ denote the observed and missing components of $X$ for a given subject, and let $R$ be the vector of observation indicators whose elements are zero or one depending on whether the corresponding element on $X$ is missing or observed. We assume throughout that the data are missing at random (MAR) \cite{Rubin:1976}. Here MAR means that $P(R|Y,X,Z)=P(R|Y,X^{obs},Z)$. We assume that $(Y_{i},X_{i},Z_{i},R_{i})$, $i=1,..,n$ are independent and identically distributed. Lastly, we let $f(Y|X,Z,\psi)$ denote the `substantive model', which is indexed by parameter $\psi$ ($\psi \in \Psi$). We assume throughout that this substantive model is correctly specified. That is, there exists $\psi \in \Psi$ such that $f_{0}(Y|X,Z)=f(Y|X,Z,\psi)$, where $f_{0}(Y|X,Z)$ denotes the true conditional distribution of $Y$ given $X$ and $Z$.

\subsection{Multiple imputation of partially observed covariates}
In Bayesian parametric MI, to multiply impute missing values in $X$ we specify a parametric model $f(X|Z,Y,\omega)$, $\omega \in \Omega$ for the conditional distribution $f(X|Y,Z)$. To create the $m$th imputed dataset we first  draw $\omega^{(m)}$ from its posterior distribution given the observed data $\{(Y,X^{obs},Z); i=1,..,n\}$ and a (usually noninformative) prior $f(\omega)$. For each subject the missing values (if any) $X^{mis}$ are imputed by taking a draw from the density $f(X^{mis}|X^{obs},Y,Z,\omega^{(m)})$ implied by $f(X|Y,Z,\omega^{(m)})$.

Having created $M$ imputed datasets, the substantive model parameter $\psi$ is then estimated separately using each imputed dataset, resulting in estimates $\hat{\psi}^{1},..,\hat{\psi}^{M}$ and corresponding variances $\widehat{\Var}(\hat{\psi}^{1}),..,\widehat{\Var}(\hat{\psi}^{M})$. In this article we assume that the substantive model is fitted using maximum likelihood. Rubin's rules are then invoked to provide a final inference for $\psi$, with the estimator of $\psi$ given by
\begin{eqnarray*}
\hat{\psi}_{MI} = \frac{\sum^{M}_{m=1} \hat{\psi}^{m}}{M},
\end{eqnarray*}
and an estimate of the variance of $\hat{\psi}_{MI}$ is given by
\begin{eqnarray*}
\widehat{\Var}(\hat{\psi}_{MI}) = \left[ \frac{1}{M} \sum^{M}_{m=1} \widehat{\Var}(\hat{\psi}^{m}) \right] + \left[ (1+1/M) \frac{1}{M-1} \sum^{M}_{m=1} (\hat{\psi}^{m} - \hat{\psi}_{MI})^{2} \right]
\label{rubinsVarianceEstimator}
\end{eqnarray*}

Suppose that the posited imputation model is correctly specified, so that there exists a value of $\omega \in \Omega$ such that $f_{0}(X|Y,Z)=f(X|Z,Y,\omega)$. Then $\hat{\psi}_{MI}$ is a consistent estimator of $\psi$, and as the number of imputations $M \rightarrow \infty$, confidence intervals based on $\hat{\Var}(\hat{\psi})$ achieve coverage at or above their nominal level \cite{Meng:1994}.

\subsection{Compatibility and imputation model mis-specification}
\label{compatibilityAndMisspec}
When an imputation model $f(X|Z,Y,\omega)$ is directly specified, it may be mis-specified if it is not \textit{compatible} with the substantive model (assuming this is correctly specified). For example, if the correctly specified substantive model includes an interaction between a partially observed covariates and a fully observed covariate in their effect on the outcome $Y$, imputation models which do not allow for this interaction will generally (unless the interaction term is in truth zero) be mis-specified. Such considerations led to recommendations that imputation models be used which do not impose restrictions which will conflict with subsequent analyses of the imputed datasets \cite{Meng:1994,Schafer:1997}.

Following Liu \etal \cite{Liu2012}, we now define the notion of compatibility between a set of conditional models. Let $A=(A_{1},..,A_{p})$ be a vector of random variables, and let $A_{-j}=(A_{1},..,A_{j-1},A_{j+1},..,A_{p})$. Then a set of conditional models $\{f_{j}(A_{j}|A_{-j},\theta_{j}); \theta_{j} \in \Theta_{j}, j=1,..,p\}$ is said to be compatible if there exists a joint model $g(A|\theta)$, $\theta \in \Theta$ and a collection of surjective maps $\{t_{j}:\Theta \rightarrow \Theta_{j}; j=1,..,p\}$ such that for each $j$, $\theta_{j} \in \Theta_{j}$, and $\theta \in t^{-1}_{j}(\theta_{j}) = \{\theta: t_{j}(\theta)=\theta_{j}\}$,
\begin{eqnarray*}
f_{j}(A_{j}|A_{-j},\theta_{j}) = g(A_{j}|A_{-j},\theta).
\end{eqnarray*}
Otherwise the set of models $\{f_{j}; j=1,..,p\}$ is said to be incompatible.

A weaker property which we shall also use is that of \textit{semi-compatibility} for a set of models. A set of models is semi-compatible if they can be made compatible by setting one or more parameters to zero. More formally (again following Liu \etal \cite{Liu2012}), a set of conditional models $\{h_{j}(A_{j}|A_{-j},\theta_{j},\kappa_{j}); \theta_{j} \in \Theta_{j}, \kappa_{j} \in K_{j}, j=1,..,p\}$ is said to be semi-compatible if there exists a set of compatible conditional models $\{f_{j}(A_{j}|A_{-j},\theta_{j}); \theta_{j} \in \Theta_{j}, j=1,..,p\}$ such that
\begin{eqnarray*}
f_{j}(A_{j}|A_{-j},\theta_{j}) = h_{j}(A_{j}|A_{-j},\theta_{j},\kappa_{j}=0),
\end{eqnarray*}
for $j=1,..,p$. A set of compatible conditional models is always semi-compatible.

Incompatibility between the imputation and substantive models does not necessarily imply mis-specification of the former. For example, suppose the substantive model is $Y|X \sim (\psi_{0}+\psi_{1}X,\sigma^{2}_{\psi})$ and the imputation model is $X|Y \sim N(\omega_{0} + \omega_{1} Y + \omega_{2} Y^{2},\sigma^{2}_{\omega})$, with each of the regression coefficients lying in $(-\infty,+\infty)$. These two models are incompatible, since there is no joint model $g(Y,X|\theta)$ satisfying the definition (this can be established by the theorem for compatibility of two conditional densities of Arnold and Press \cite{Arnold1989}). The models are semi-compatible however, with corresponding joint model $g(X,Y)$ equal to the bivariate normal model, by setting $\omega_{2}=0$. Despite incompatibility, the imputation model is not necessarily mis-specified: for example if $(X,Y)$ is in truth bivariate normal, $\hat{\psi}_{MI}$ is consistent for $\psi$. Here incompatibility does not imply mis-specification because a more restrictive version of the imputation model (with $\omega_{2}=0$) is compatible with the substantive model.

Conversely, suppose the substantive model is $Y|X \sim N(\psi_{0}+\psi_{1} X + \psi_{2} X^{2},\sigma^{2}_{\psi})$ and the imputation model is $X|Y \sim N(\omega_{0} + \omega_{1} Y,\sigma^{2}_{\omega})$. Then again the models are semi-compatible but not compatible, and unless $\psi_{2}=0$ in truth, the imputation model will be mis-specified, since there exist no joint model with conditionals corresponding to the substantive and imputation models. Consequently the MI estimator $\hat{\psi}_{MI}$ will be inconsistent, as demonstrated through simulation by von Hippel \cite{Hippel2009} and Seaman \etal \cite{Seaman2012}.

Incompatibility may also arise when default imputation models are used for covariates in non-linear substantive models. For example, suppose $T$ (rather than $Y$) is a time to event outcome which is not subject to censoring, and that the substantive model is the parametric exponential model, with hazard function $h(t)=h_{0} \exp(\psi X)$, with $X$ a continuous partially observed covariate. In this case $H_{0}(T)=\int^{T}_{0} h_{0}(t) dt = h_{0}T$. Then suppose, following the recent recommendations of White and Royston \cite{White2009}, we adopt a normal linear imputation model for $X|T \sim N(\omega_{0}+\omega_{1}T,\sigma^{2}_{\omega})$ with $T \propto H_{0}(T)$ as covariate. Then the two models are incompatible, and consequently the MI estimator $\hat{\psi}_{MI}$ is inconsistent (although simulations by White and Royston \cite{White2009} show the bias is often small).

In conclusion, except in cases where the imputation and substantive models can be made compatible by restricting the parameter space $\Omega$ of the imputation model, incompatibility between the two implies the imputation model is mis-specified (assuming correct specification of the substantive model). Consequently, when choosing the covariate imputation model $f(X|Z,Y,\omega)$ we should (at least) ensure that it is either compatible with the substantive model, or a restriction of it is compatible with the substantive model.

\subsection{Joint model imputation}
\label{jointmodelMI}
The natural route to ensuring compatibility between the imputation and substantive models is to explicitly specify a joint model $g(Y,X|Z,\theta)$ which has the substantive model $f(Y|X,Z,\psi)$ as its corresponding conditional, and to derive the implied imputation model. Given the (correctly) specified substantive model $f(Y|X,Z,\psi)$, such joint models are specified by defining a model $f(X|Z,\delta)$. The imputation model is then given by
\begin{eqnarray}
f(X|Z,Y,\psi,\delta) = \frac{f(Y,X|Z,\psi,\delta)}{f(Y|Z,\psi,\delta)} &=& \frac{f(Y|X,Z,\psi) f(X|Z,\delta)}{f(Y|Z,\psi,\delta)} \nonumber \\
& \propto & f(Y|X,Z,\psi) f(X|Z,\delta).
\label{imputationModel}
\end{eqnarray}
We emphasize that using a compatible imputation model does not guarantee it is correctly specified - this is only true if, in addition to the substantive model being correctly specified, $f(X|Z,\delta)$ is correctly specified.

In cases where $p=1$ and $X$ is univariate, specification of a model $f(X|Z,\delta)$ is relatively straightforward. When $X$ is multivariate, and particularly when it contains a mixture of continuous and discrete variables, specification of a joint model $f(X|Z,\delta)$ becomes more challenging. In this setting Ibrahim \etal \cite{Ibrahim:1999} proposed specification by factorising the joint distribution of $X|Z$ as a product of univariate densities of the form
\begin{eqnarray}
f(X_{1}|Z) f(X_{2}|X_{1},Z) f(X_{3}|X_{1},X_{2},Z) \ldots.
\label{ibrahimFactorisation}
\end{eqnarray}
This breaks the problem of joint specification into the easier task of specification of a series of univariate models. This means that appropriate univariate regression models can be specified depending on the type (i.e. continuous, discrete, ordered discrete) of each variable. However, as the dimension of $X$ increases the number of possible orderings of its components increases rapidly, and it is not obvious which ordering should be chosen. As far as we aware this approach to MI has not been adopted by applied researchers.

\subsection{Fully conditional specification imputation}
In the more general setting of MI in multivariate data, the fully conditional specification (FCS) approach to MI, which we describe in detail in the following section, similarly splits the task of specification of a joint model into a series of univariate model specifications. In FCS MI models are specified for each partially observed variable conditional on all other variables. In contrast to the approach proposed by Ibrahim \etal \cite{Ibrahim:1999}, no choice of ordering for model specification is required. FCS MI has now become an extremely popular approach to MI generally \cite{Buuren:2007}.

Application of FCS for imputation of partially observed covariates involves specification of models of the form $f(X_{j}|X_{-j},Z,Y)$, where $X_{-j}$ denotes the components of $X$ with $X_{j}$ removed. As we have described, for certain substantive models, such as those involving non-linear covariate effects or interactions, default choices of these imputation models within FCS will be incompatible (and further, not semi-compatible) with the substantive model, and will therefore be mis-specified. Motivated by this, in Section \ref{accommodatingTheOutcomeModel} we propose a modification of FCS MI which ensures that each of the covariate models $f(X_{j}|X_{-j},Z,Y)$ is compatible with the substantive model.

\section{Review of fully conditional specification multiple imputation}
\label{fcs}
In this section we review the fully conditional specification (FCS) approach to MI \cite{Buuren:2007}, within the setup of partially observed covariates previously defined in Section \ref{compatibility}.

\subsection{The fully conditional specification algorithm}
For each partially observed covariate $X_{j}$, we posit an imputation model $f(X_{j}|X_{-j},Z,Y,\theta_{j})$, with parameter $\theta_{j}$, where $X_{-j}=(X_{1},..,X_{j-1},X_{j+1},..,X_{p})$. This is typically a generalised linear model chosen according to the type of $X_{j}$ (e.g. continuous, binary, count). Furthermore, a non-informative prior distribution $p(\theta_{j})$ for $\theta_{j}$ is specified. Let $x^{obs}_{j}$ and $x^{mis}_{j}$ denote the vectors of observed and missing values in $X_{j}$ for the $n$ subjects. Let $y$ and $z$ denote the vector and matrix of (fully observed) values of $Y$ and $Z$ across the $n$ subjects.

The FCS algorithm begins by replacing the missing values in each $X_{j}$ by randomly selected observed values from the same variable. The algorithm then proceeds by repeatedly imputing the missing values in each variable, at each stage conditioning on the most recent imputations of the other variables. Let $x^{mis(t)}_{j}$ denote the imputations of the missing values $x^{mis}_{j}$ at iteration $t$ and let $x^{(t)}_{j}=(x^{obs}_{j},x^{mis(t)}_{j})$ denote the vector of observed and imputed values at iteration $t$. The $t$th iteration of the algorithm consists of drawing from the following distributions (up to constants of proportionality):
\begin{equation}
\left.
\begin{aligned}
\theta^{(t)}_{1} &\sim p(\theta_{1}) f(x^{obs}_{1}|x^{(t-1)}_{2},..,x^{(t-1)}_{p},z,y, \theta_{1}) \\
x^{mis(t)}_{1} &\sim f(x^{mis}_{1}|x^{(t-1)}_{2},..,x^{(t-1)}_{p},z,y, \theta^{(t)}_{1})  \\
\theta^{(t)}_{2} &\sim  p(\theta_{2}) f(x^{obs}_{2}|x^{(t)}_{1},x^{(t-1)}_{3},..,x^{(t-1)}_{p},z,y,\theta_{2})  \\
x^{mis(t)}_{2} &\sim f(x^{mis}_{2}|x^{(t)}_{1},x^{(t-1)}_{3},..,x^{(t-1)}_{p},z,y, \theta^{(t)}_{2})  \\
\vdots  \\
\theta^{(t)}_{p} &\sim  p(\theta_{p}) f(x^{obs}_{p}|x^{(t)}_{1},..,x^{(t)}_{p-1},z,y, \theta_{p})  \\
x^{mis(t)}_{p} &\sim f(x^{mis}_{p}|x^{(t)}_{1},..,x^{(t)}_{p-1},z,y, \theta^{(t)}_{p})  \\
\end{aligned}
\right\}
\label{fcsEquations}
\end{equation}
Thus, for the partially observed covariate $X_{j}$ the algorithm first draws from the posterior distribution of the corresponding imputation model parameters determined by the prior and likelihood corresponding to the imputation model fitted to data from subjects for whom $X_{j}$ was observed, conditional on all the other variables (observed plus most recently imputed values). Note that in this respect the FCS algorithm differs from a standard Gibbs sampler, which at the same step would additionally condition on $x^{mis}_{j}$ from the previous iteration. Missing values in $X_{j}$ are then imputed from the imputation model using the parameter value drawn in the preceding step. After a sufficient number of iterations it is assumed that the algorithm has converged to a stationary distribution, and the final draws of the missing data form a single imputed dataset. The process is then repeated to create as many imputed datasets as desired. In software implementations of FCS MI the variables are typically updated in the ordering for which the missingness pattern is closest to monotone. Finally, the substantive model is fitted to each imputed dataset, and the results combined using Rubin's Rules as described previously.

\subsection{Statistical properties}
\label{fcsProperties}
Despite the fact FCS MI has been applied widely in a number of fields \cite{Buuren:2007}, until recently few results were available regarding its validity. This is due to the fact that one can specify imputation models (which in our setting are $f(X_{j}|X_{-j},Z,Y,\theta_{j})$) that are not mutually compatible (\cite{vanBuuren2006}). In this case it is not clear to what distribution, if any, the algorithm will converge.

Liu \etal have recently given sufficient conditions under which FCS MI is asymptotically equivalent to MI from a Bayesian joint model \cite{Liu2012}. Principal among these is that the set of conditional imputation models are mutually compatible. However, compatibility is not sufficient for equivalence between FCS and MI from a Bayesian joint model. One situation in which equivalence does not hold despite the imputation models being compatible is when information regarding parameters of a conditional model is contained, but not utilised, in the marginal distribution of the variables being conditioned on. An example of this is when a binary variable is imputed using logistic regression conditional on a continuous variable, with the latter imputed using a normal linear regression model. Although these models are compatible with each other, FCS imputation fails to utilise the information in the conditional distribution of the continuous variable given the binary variable regarding the logistic regression parameters, and consequently FCS MI is not equivalent to MI from a Bayesian model.

Liu \etal further show that provided each of the conditional models is correctly specified, the estimator $\hat{\psi}_{MI}$ is consistent \cite{Liu2012}. Thus if a set of conditional models is used which are semi-compatible, and each is correctly specified, consistent estimates are obtained. Note that in this case since FCS is not necessarily equivalent to imputation from a Bayesian joint model, there is no guarantee that Rubin's rule for the variance will provide valid inferences. This result can also be used to conclude that in the linear-logistic example described in the previous paragraph, in which the models are compatible (and therefore also semi-compatible) $\hat{\psi}_{MI}$ is consistent provided both models are correctly specified. If the conditional models used are not even semi-compatible, in general we expect inconsistent estimates.

\section{Fully conditional specification imputation accommodating the substantive model}
\label{accommodatingTheOutcomeModel}
In this section we describe how the standard FCS algorithm, described in Section \ref{fcs}, can be modified to ensure that each of the univariate imputation models used is compatible with the substantive model. We term the algorithm substantive model compatible FCS (SMC-FCS).

\subsection{The algorithm}
First we specify a non-informative prior for the parameters of the substantive model, denoted $f(\psi)$. Then for each $j=1,..,p$ we specify a model $f(X_{j}|X_{-j},Z,\phi_{j})$ ($j=1,..,p$) and non-informative prior $f(\phi_{j})$. At iteration $t$, for $j=1,..,p$ we first draw $\psi^{(t,j)}$ from
\begin{eqnarray}
\psi^{(t,j)} & \sim  f(\psi) f(y|x^{(t)}_{1},..x^{(t)}_{j-1},x^{(t-1)}_{j},..,x^{(t-1)}_{p},z,\psi).
\label{substantiveModelPosterior}
\end{eqnarray}
Then a draw $\phi^{(t)}_{j}$ is made from
\begin{eqnarray}
\phi^{(t)}_{j} &\sim   f(\phi_{j}) f(x^{(t-1)}_{j} | x^{(t)}_{1},..,x^{(t)}_{j-1},x^{(t-1)}_{j+1},..,x^{(t-1)}_{p},z,\phi_{j}).
\label{covariateModelPosterior}
\end{eqnarray}
Note that here, as in a standard Gibbs sampler, the draw is made from the posterior corresponding to the fit of the model $f(X_{j}|X_{-j},Z,\phi_{j})$ using data (imputed and observed) from all subjects, rather than to only those subjects for whom $X_{j}$ is observed. This is necessary since, if missingness in $X_{j}$ is dependent on $Y$, drawing from the posterior corresponding to the model fitted to only those with $X_{j}$ observed would introduce bias.

For each subject with $X_{j}$ missing we then impute their missing value from the density proportional to
\begin{eqnarray}
f(Y|X^{(t)}_{1},..,X^{(t)}_{j-1},X_{j},X^{(t-1)}_{j+1},..,X^{(j-1)}_{p},Z,\psi^{(t,j)}) f(X_{j}|X^{(t)}_{1},..,X^{(t)}_{j-1},X^{(t-1)}_{j+1},..,X^{(j-1)}_{p},Z,\phi^{(t)}_{j}).
\label{targetdistribution}
\end{eqnarray}
By construction, following equation \eqref{imputationModel}, this density will be compatible with the substantive model $f(Y|X,Z,\psi)$. However, in general the density will not belong to a standard parametric family, complicating direct simulation of values. In Section \ref{rejectionSamplingSubsection} we therefore show how rejection sampling can be used to draw from the density, giving implementations for a number of important types of substantive model.

\subsection{Statistical properties}
SMC-FCS is an example of a (possibly incompatible) Gibbs sampler. However, as with standard FCS, determining the statistical properties of the algorithm in generality is challenging. In some special cases the algorithm corresponds to a Gibbs sampler for a well defined Bayesian model. This will be true when there exists a joint model $g(X|Z,\gamma)$ and prior $f(\gamma)$ for which the conditionals required for a Gibbs sampler correspond to those in equations \eqref{substantiveModelPosterior}, \eqref{covariateModelPosterior} and \eqref{targetdistribution}. Since in this case SMC-FCS is equivalent to imputation from a Bayesian joint model, Rubin's rules can then be applied for inference.

As for standard FCS MI, compatibility between the models $f(X_{j}|X_{-j},Z,\phi_{j})$ is not sufficient for equivalence with Bayesian MI. If one partially observed covariate is modelled conditional on a second using logistic regression, with the second modelled with normal linear regression given the first, SMC-FCS is not equivalent to Bayesian joint model MI, despite these models being compatible, for the reason given in Section \ref{fcsProperties}.

If the covariate models $f(X_{j}|X_{-j},Z,\phi_{j})$ are semi-compatible and correctly specified, we conjecture that, the MI estimator of the substantive model parameters $\hat{\psi}_{MI}$ will be consistent. We investigate this in Section \ref{simulationSection} using simulations. In cases where SMC-FCS is consistent but not equivalent to imputation from a Bayesian joint model there is no guarantee that confidence intervals based on Rubin's variance estimator will give at least nominal coverage. Lastly, if the covariate models are not semi-compatible, in general we do not expect the estimator $\hat{\psi}_{MI}$ to be consistent.

In software implementations of the standard FCS algorithm 10 iterations are typically used to `burn-in', based on empirical experience suggesting this is often sufficient for convergence of the sampler. Since when imputing missing values in $X_{j}$ our proposed modification of FCS involves fitting models using the most recently imputed values of $X_{j}$, we expect SMC-FCS to require a larger number of iterations in order to converge to the required stationary distribution, assuming it exists.

\section{Sampling from the imputation model}
\label{rejectionSamplingSubsection}
In this section we give details of how the method of rejection sampling can be used to sample from the density given in equation \eqref{targetdistribution} for some of the most common types of substantive model. Rejection sampling involves creating draws from a proposal density (from which it is easy to draw), until a draw is made satisfying a particular condition. Suppressing the iteration index $t$, we choose $f(X_{j}|X_{-j},Z,\phi_{j})$ as our proposal density, on the assumption that it is easy to sample from this density. To use rejection sampling, the ratio of the target density to the proposal density (up to a constant of proportionality) must be bounded  above in $X_{j}$ \cite{Gelman2004}. Here this ratio is proportional to:
\begin{eqnarray}
\frac{f(Y|X,Z,\psi)f(X_{j}|X_{-j},Z,\phi_{j})}{f(X_{j}|X_{-j},Z,\phi_{j})} = f(Y|X_{j},X_{-j},Z,\psi).
\label{ratioofdensities}
\end{eqnarray}
Let $c(Y,X_{-j},Z,\psi)$ denote an upper bound (in $X_{j}$) for $f(Y|X_{j},X_{-j},Z,\psi)$. To generate a draw from the density proportional to equation \eqref{targetdistribution}, we sample pairs of values $X^{*}_{j}$ from the density given by $f(X_{j}|X_{-j},Z,\phi_{j})$ and $U$ from the uniform distribution on (0,1) until:
\begin{eqnarray}
U \leq \frac{f(Y|X^{*}_{j},X_{-j},Z,\psi)}{c(Y,X_{-j},Z,\psi)}.
\label{rjcriterion}
\end{eqnarray}
When this inequality is satisfied, the value $X^{*}_{j}$ is a draw from the density proportional to equation \eqref{targetdistribution}.

We must therefore bound $f(Y|X_{j},X_{-j},Z,\psi)$ in $X_{j}$. The bound will depend on the specification of the substantive model. In the following we derive bounds for the cases of i) a normal regression model, ii) a model for a discrete outcome $Y$, and iii) a proportional hazards survival model.

\subsection{Normal regression}
Suppose that the substantive model specifies that $Y$ is normal, with conditional mean $E(Y|X)=g(X_{j},X_{-j},Z,\beta)$ for some function $g()$, and residual variance $\sigma^{2}_{\epsilon}$, so that $\psi=(\beta,\sigma^{2}_{\epsilon})$. Then:
\begin{eqnarray*}
f(Y|X_{j},X_{-j},Z,\psi) &=& \frac{1}{\sqrt{2\pi \sigma^{2}_{\epsilon}}} \exp(-(Y-g(X_{j},X_{-j},Z,\beta))^{2}/2\sigma^{2}_{\epsilon}) \\
& \leq & \frac{1}{\sqrt{2\pi \sigma^{2}_{\epsilon}}}.
\end{eqnarray*}
To generate a draw for the missing value $X_{j}$, we draw a value $X^{*}_{j}$ from $f(X_{j}|X_{-j},Z,\phi_{j})$, and $U$ from the uniform distribution on $(0,1)$. The draw $X^{*}_{j}$ is accepted if:
\begin{eqnarray}
U &\leq&  f(Y|X^{*}_{j},X_{-j},Z,\psi) \sqrt{2\pi \sigma^{2}_{\epsilon}}  \nonumber \\
&=& \exp(-(Y-g(X^{*}_{j},X_{-j},Z,\beta))^{2}/2\sigma^{2}_{\epsilon}) .
\label{eqrejectioncondition}
\end{eqnarray}
If the draw is not accepted, new draws of $X^{*}_{j}$ and $U$ are made until they satisfy the condition in equation \eqref{eqrejectioncondition}.

\subsection{Discrete outcomes}
Now consider a discrete outcome $Y$. This includes the case of a binary outcome $Y$, which is commonly modelled using logistic regression. When $Y$ is discrete, $f(Y|X_{j},X_{-j},Z,\psi)$ is a probability, and hence is less than or equal to one. The rejection sampling algorithm then consists of drawing $X^{*}_{j}$ from $f(X_{j}|X_{-j},Z,\phi_{j})$ and $U \sim U(0,1)$, and accepting $X^{*}_{j}$ when:
\begin{eqnarray*}
U &\leq& f(Y|X^{*}_{j},X_{-j},Z,\psi).
\end{eqnarray*}

\subsection{Proportional hazards models}
\label{proportionalHazardsModels}
Lastly, suppose that interest lies in the time $T$ to an event of interest, but this time may be censored. Let $C$ denote the censoring time. We observe $W=\mbox{min}(T,C)$, and $D=1(T<C)$, which denotes whether the subject's event time has been observed or was censored. We assume that censoring is noninformative, in the sense that $T \cind C | X,Z$. Furthermore, we assume $C \cind X | Z$. Together these assumptions allow us to avoid modelling the censoring process (\cite{Rathouz2007}).

We assume that the substantive model is the proportional hazards model:
\begin{eqnarray}
h(t|X)=h_{0}(t) \exp(g(X_{j},X_{-j},Z,\beta))
\end{eqnarray}
where $h(t|X)$ denotes the hazard at time $t$, $h_{0}(t)$ denotes the baseline hazard at time $t$ and $g(X_{j},X_{-j},Z,\beta)$ denotes a function of $X$ and $Z$ indexed by parameter $\beta$. In parametric proportional hazards models the baseline hazard function is parametrized by a finite set of parameters $\lambda$, so that $\psi=(\beta,\lambda)$. In Cox's proportional hazards model the baseline hazard is allowed to be arbitrary, so that $\psi=(\beta,h_{0}(.))$ with $h_{0}(.)$ an infinite dimensional parameter. Equivalently we can parametrize the model using the baseline cumulative hazard $H_{0}(t)=\int^{t}_{0} h_{0}(s) ds$, so that $\psi=(\beta,H_{0}(.))$.

We first consider how to sample $X_{j}$ for a subject for whom $D=0$, i.e. their time to event is censored. Then since by assumption $T \cind C | X,Z$ and $C \cind X | Z$,
\begin{eqnarray*}
f(W=t,D=0|X_{j},X_{-j},Z,\psi) &=& f(T>t, C=t | X_{j}, X_{-j},Z,\psi) \\
&=& P(T>t | X_{j}, X_{-j},Z,\psi) f(C=t|X_{j},X_{-j},Z) \\
&=& P(T>t | X_{j}, X_{-j},Z,\psi) f(C=t|Z) \\
& \leq & f(C=t|Z).
\end{eqnarray*}
We draw $X^{*}_{j}$ from $f(X_{j}|X_{-j},Z,\phi_{j})$ and $U \sim U(0,1)$, and accept $X^{*}_{j}$ when:
\begin{eqnarray*}
U &\leq& \frac{f(W=t,D=0|X^{*}_{j},X_{-j},Z,\psi)}{f(C=t|Z)} \\
&=& P(T>t|X^{*}_{j},X_{-j},Z,\psi) \\
&=& \exp(-H_{0}(t) e^{g(X^{*}_{j},X_{-j},Z,\beta)}),
\end{eqnarray*}
where $H_{0}(t)=\int^{t}_{0} h_{0}(s) ds$ denotes the cumulative baseline hazard function.

For a subject who is not censored ($D=1$), we have:
\begin{eqnarray*}
f(W=t,D=1|X_{j},X_{-j},Z,\psi) &=& P(C>t|X_{j}, X_{-j},Z) h(t|X_{j},X_{-j},Z,\psi) P(T>t|X_{j},X_{-j},Z,\psi) \\
&=& P(C>t|Z) h_{0}(t) \exp(g(X_{j},X_{-j},Z,\beta)-H_{0}(t) e^{g(X_{j},X_{-j},Z,\beta)}).
\end{eqnarray*}
Since $\exp()$ is monotonically increasing, this expression takes its maximum when $g(X_{j},X_{-j},\beta)-H_{0}(t) e^{g(X_{j},X_{-j},\beta)}$ takes its maximum. Differentiating this with respect to $g()$ and setting the resulting expression to zero shows that this occurs when $H_{0}(t) e^{g(X_{j},X_{-j},\beta)}=1$. Therefore:
\begin{eqnarray*}
f(W=t,D=1|X_{j},X_{-j},Z,\psi) \leq P(C>t|Z) \frac{h_{0}(t) e^{-1}}{H_{0}(t)}.
\end{eqnarray*}
We can thus draw $X^{*}_{j}$ from $f(X_{j}|X_{-j},Z,\phi_{j})$ and $U \sim U(0,1)$, and accept $X^{*}_{j}$ when:
\begin{eqnarray*}
U &\leq& \frac{f(W=t,D=1|X^{*}_{j},X_{-j},Z,\psi)}{P(C>t|Z) \frac{h_{0}(t) e^{-1}}{H_{0}(t)}} \\
&=& \exp(1+g(X^{*}_{j},X_{-j},Z,\beta)-H_{0}(t) e^{g(X^{*}_{j},X_{-j},Z,\beta)}) H_{0}(t).
\end{eqnarray*}

\section{Simulation study}
\label{simulationSection}
In this section we describe the results of simulation studies to investigate the performance SMC-FCS in situations in which the substantive model is incompatible with standard choices for covariate imputation models.

\subsection{Linear regression with quadratic covariate effects}
\label{quadraticSimulations}
We first simulated from a linear regression substantive model with a single covariate $X$ with linear and quadratic effects, for which standard imputation model choices for the covariate $X$ are incompatible.

\subsubsection{Simulation setup}
The outcome $Y$ was simulated according to:
\begin{eqnarray*}
Y = \beta_{0} + \beta_{1} X + \beta_{2} X^{2} + \epsilon,
\end{eqnarray*}
with $\beta_{0}=4$, $\beta_{1}=-4$, $\beta_{2}=1$ and $\epsilon \iid N(0,\sigma^{2}_{\epsilon})$. These coefficients were chosen to give a moderately strong U-shaped association between $Y$ and $X$. The variance $\sigma^{2}_{\epsilon}$ was chosen such that the coefficient of determination $R^{2}$ was equal to 0.5.

The covariate $X$ was simulated from a normal, a log-normal, or a normal mixture distribution. For all three distributions $X$ had mean 2 and variance 1. For the log-normal distribution, $X$ was generated by exponentiating a draw from $N(\log(\sqrt{3.2}),\log{(5/4)})$. For the normal mixture distribution, $X$ was drawn from $N(1.125, 0.234)$ with probability 0.5 and from $N(2.875, 0.234)$ with probability 0.5.

For each distribution of $X$, values were made missing either according to the MCAR mechanism $P(R=1|X,Y)=0.7$ or the MAR mechanism $P(R=1|X,Y)=\expit(\alpha_{0} + \alpha_{1} Y)$, where $\expit(a)=(1+\exp(-a))^{-1}$, $\alpha_{1}=-1/\mbox{SD}(Y)$ and $\alpha_{0}$ was chosen to make the marginal probability of observing $X$ equal to 0.7. In all simulations datasets for $n=1,000$ subjects were generated, and 1,000 simulations were performed for each scenario.

\subsubsection{Estimation methods}
For each simulated dataset we first imputed the missing values of $X$ using a linear regression model with the {\tt ice} command in Stata. We used the default imputation model, with the expectation of $X$ modelled as a linear function of $Y$. Note that since here there is only one partially observed variable, no iteration is required within FCS. Missing values of $X^{2}$ were then passively imputed as the square of these imputed values of $X$ (`linear imputation'). Second, we imputed the missing $X$ values using the `transform then impute' or `just another variable' (JAV) approach proposed by \cite{Hippel2009}, that is, by treating $X^{2}$ as another variable to be imputed in the {\tt ice} command in Stata. Third, we imputed $X$ using the {\tt mice.impute.quadratic} function in the R MICE package (`polynomial combination'). This implements a method recently proposed by Van Buuren (p140 \cite{vanBuuren2012}), which imputes the linear combination of $X$ and $X^2$ which enters in the linear predictor of the substantive model, followed by solving a quadratic equation for $X$. Lastly, we used SMC-FCS, assuming $X$ is marginally normally distributed for all scenarios. We chose to implement SMC-FCS using the same marginal model for $X$ to explore the performance of (substantive model) compatible but mis-specified imputation models. For all imputation approaches, 10 imputed datasets were generated, and estimates and confidence intervals (CI) found using Rubin's rules. We used 10 iterations per imputation in SMC-FCS, and the default 10 iterations in the {\tt ice} command. With $X$ univariate, SMC-FCS is equivalent to imputation from the corresponding Bayesian model. We used standard non-informative priors for normal linear regression parameters in SMC-FCS, i.e. $f(\beta, \sigma^{2}) \propto 1/\sigma^{2}$, where $\beta$ and $\sigma^{2}$ denote the vector of regression coefficients and residual variance respectively.

\subsubsection{Results}
Table \ref{linearRegQuadraticResults} shows the results of the simulations, showing the empirical mean and standard deviation of estimates of $\beta_{2}$ and the coverage of nominal 95\% confidence intervals. With normally distributed $X$ and MCAR, linear imputation resulted in biased estimates, with considerable attenuation in $\hat{\beta}_{2}$ towards zero as expected. Here the imputation model being used is incompatible (with the substantive model) and mis-specified. Confidence interval coverage for linear imputation was also extremely poor, with zero coverage for $\beta_{2}$. JAV, polynomial combination and SMC-FCS gave unbiased results, with similar efficiency to each other. JAV and SMC-FCS had CI coverage close to 95\%, but polynomial combination had slightly low coverage. With $X$ log-normally distributed and MCAR linear imputation was again biased with poor CI coverage. JAV was unbiased, although estimates were considerably more variable SMC-FCS. Furthermore, the coverage of the CI for $\beta_{2}$ from JAV was only 84\%. The polynomial combination method performed similarly to JAV here. Despite the assumed model for $X$ being mis-specified, SMC-FCS was unbiased and the 95\% CI for $\beta_{2}$ had the correct coverage. With $X$ distributed according to a normal mixture model and MCAR, JAV and polynomial combination were again unbiased. The CI coverage of JAV and polynomial combination for $\beta_{2}$ was close to 95\%. Linear imputation continued to be severly biased. SMC-FCS was somewhat biased towards the null for $\beta_{2}$, and consequently CI coverage for $\beta_{2}$ was only 74\%.

With normal $X$ and MAR, linear imputation gave biased estimates and the CI for $\beta_{2}$ had zero coverage. With data MAR, JAV no longer gave unbiased estimates, in agreement with the findings of \cite{Seaman2012}, and the CI for $\beta_{2}$ had only 18\% coverage. Polynomial combination had only slight bias, but CI coverage for $\beta_{2}$ was only 75.9\%. In contrast, SMC-FCS was unbiased and the CI for $\beta_{2}$ had approximately 95\% coverage. All estimators were considerably more variable with $X$ log-normal MAR. JAV and polynomial combination had considerable bias and poor CI coverage for $\beta_{2}$, as did linear imputation. Despite using a mis-specified model for the distribution of $X$, SMC-FCS was again unbiased, although the CI for $\beta_{2}$ had somewhat lower than nominal coverage. Lastly with $X$ distributed as a mixture of two normals and MAR, linear imputation continued to be substantially biased. JAV was biased to a lesser extent, although its CI for $\beta_{2}$ had poor coverage. SMC-FCS was biased (since the assumed distribution for $X$ was incorrect) towards zero, and its CI for $\beta_{2}$ had extremely poor coverage. The polynomial combination method performed best here, with unbiased estimates of $\beta_{2}$ and only somewhat reduced CI coverage.

\begin{table}
\caption{\label{linearRegQuadraticResults}Simulation results - linear regression with quadratic covariate effects. Empirical mean (SD) of estimates of quadratic coefficient $\beta_{2}=1$ from 1,000 simulations, using linear imputation (linear), just another variable imputation (JAV), the polynomial combination method, and SMC-FCS. Empirical coverage of nominal 95\% confidence intervals is also shown (Cov). Monte-Carlo errors for means and SDs are less than 0.003, except for log-normal $X$ MAR, where Monte-Carlo errors for means and SDs are less than 0.02. Monte-Carlo errors for confidence interval coverage are less than 1.6\%.}
\centering
\fbox{%
\begin{tabular}{lllllllll}
Scenario & \multicolumn{2}{c}{Linear} & \multicolumn{2}{c}{JAV} & \multicolumn{2}{c}{Polynomial comb.} & \multicolumn{2}{c}{SMC-FCS} \\
& Mean (SD) & Cov & Mean (SD) & Cov & Mean (SD) & Cov & Mean (SD) & Cov \\
\hline
$X$ MCAR \\
Normal $X$   &  0.693 (0.040) & 0.0 & 0.999 (0.041) & 93.8 & 1.005 (0.040) & 91.5 & 0.998 (0.038) & 94.7 \\
Log-normal $X$  &   0.789 (0.085) & 18.0 & 1.001 (0.099) & 83.8 & 1.025 (0.097) & 83.1 &  1.000 (0.059) & 95.6 \\
$X$ mixture of normals  &  0.489 (0.035) & 0.0 & 0.995 (0.036) & 95.9 & 1.003 (0.034) & 95.8 & 0.942 (0.036) & 65.2 \\
\hline
$X$ MAR \\
Normal $X$ &  0.610 (0.045) & 0.0 & 1.186 (0.074) & 18.0 & 1.045 (0.069) & 75.9 &  0.994 (0.049) & 93.5 \\
Log-normal $X$ &   0.786 (0.275) & 53.4 & 1.462 (0.322) & 33.5 & 1.288 (0.179) & 27.6 & 1.007 (0.159) & 90.6 \\
$X$ mixture of normals &  0.443 (0.033) & 0.0 & 1.081 (0.047) & 58.8 & 1.009 (0.048) & 87.8 &  0.841 (0.037) & 2.7 \\
\end{tabular}
}
\end{table}

\subsection{Linear regression with interaction}
\label{linRegInteraction}
Next we considered a linear regression substantive model in which two covariates interact with each other in their effect on outcome.

\subsubsection{Simulation setup}
The outcome $Y$ was generated according to:
\begin{eqnarray*}
Y = \beta_{0} + \beta_{1} X_{1} + \beta_{2} X_{2} + \beta_{3} X_{1} X_{2} + \epsilon,
\end{eqnarray*}
with $\beta_{0}=0$, $\beta_{1}=1$, $\beta_{2}=1$, $\beta_{3}=1$, and with $\epsilon \iid N(0,\sigma^{2}_{\epsilon})$, where as before, $\sigma^{2}_{\epsilon}$ was chosen to give $R^{2}=0.5$.

In the first scenario $X_{1}$ and $X_{2}$ were generated from a bivariate normal distribution, with each covariate having mean 2 and variance 1, and the correlation between the two equal to 0.5. To explore robustness of the imputation methods to violations of normality assumptions, in a second scenario $\log(X_{1})$ and $\log(X_{2})$ were generated from a bivariate normal distribution so that they both had marginal distribution $N(\log(\sqrt{3.2}),\log{5/4})$ and the correlation between the two was equal to 0.5. To investigate robustness to linearity assumptions between covariates, in a third scenario we generated $X_{1} \sim N(2,1)$ and $X_{2}|X_{1} \sim N((X_{1}-2)^{2},2)$. Fourth, we generated $X_{1}$ from a Bernoulli distribution with probability 0.5, and $X_{2}|X_{1} \sim N(X_{1},1)$. To explore robustness to violations of normality assumptions, in the final scenario we generated $X_{1}$ from a Bernoulli distribution with probability 0.5 and $X_{2} = X_{1} + \exp(v)$ where $v \sim N(\log(\sqrt{3.2}),\log{(5/4)})$.

Values in both $X_{1}$ and $X_{2}$ were first made (independently) MCAR, each with probability 0.7 of being observed. We then repeated the simulations with $X_{1}$ observed with probability $\expit(\alpha_{0} + \alpha_{1} Y)$ where $\alpha_{1}=-1/\mbox{SD}(Y)$ and $\alpha_{0}$ was chosen to make the marginal probability of observing $X$ equal to 0.7, and with $X_{2}$ also observed with probability $\expit(\alpha_{0} + \alpha_{1} Y)$.

\subsubsection{Estimation methods}
For each simulated dataset, as before, estimates were obtained first using CC. In `FCS' missing values in $X_{1}$ and $X_{2}$ were imputed using the {\tt ice} command in Stata. A linear regression imputation model was used when the covariate was continuous and a logistic regression imputation model when the covariate was binary. In the imputation model for $X_{1}$ ($X_{2}$) the outcome $Y$, $X_{2}$ ($X_{1}$) and their interaction $YX_{2}$ ($YX_{1}$) were included as explanatory variables.

We obtained estimates using JAV by including the interaction variable $X_{1}X_{2}$ as an additional variable in the {\tt ice} command. Covariate $X_{1}$ ($X_{2}$) was imputed using a linear regression model (even when $X_{1}$ was binary) with $Y$, $X_{2}$ ($X_{1}$) and $X_{1}X_{2}$ as explanatory variables. The interaction term $X_{1}X_{2}$ was imputed using a linear regression model with $Y$, $X_{1}$ and $X_{2}$ as explanatory variables. Since all conditional imputation models are linear regressions with other variables included linearly, FCS is here equivalent to imputation from a trivariate normal imputation model for $(X_{1},X_{2},X_{1}X_{2})$ conditional on $Y$.

Lastly, we obtained estimates using SMC-FCS, assuming a normal regression model for $X_{1}|X_{2}$ or a logistic regression model when $X_{1}$ was binary. A linear regression model was assumed for $X_{2}|X_{1}$. When assuming $X_{1}|X_{2}$ and $X_{2}|X_{1}$ are linear regressions, SMC-FCS is equivalent to imputation from the Bayesian model defined by the substantive model and a bivariate normal model for $(X_{1},X_{2})$. In contrast, when assuming a logistic regression model for $X_{1}|X_{2}$ and a linear regression for $X_{2}|X_{1}$, although these models are compatible, SMC-FCS is not equivalent to imputation from a Bayesian model. When drawing from the posterior of the logistic regression parameters (equation \eqref{covariateModelPosterior}) we used a multivariate normal, with mean equal to the MLE and variance covariance corresponding to the inverse of the `observed' data information matrix.

\subsubsection{Results}
Table \ref{linearRegInteractionResults} shows the mean (SD) estimates of $\beta_{1}$ and $\beta_{3}$ and empirical coverage of the corresponding 95\% confidence intervals. With data MCAR, CC was unbiased as expected. With $X_{1}$ and $X_{2}$ bivariate normal, FCS imputation was substantially biased and had poor CI coverage for $\beta_{0}$ and $\beta_{3}$. In contrast both JAV and SMC-FCS were unbiased. However, SMC-FCS was somewhat more efficient than CC and JAV. Confidence interval coverage for $\beta_{3}$ was at the nominal level for both JAV and congenial FCS. With $X_{1}$ and $X_{2}$ distributed log-normal, FCS had slightly larger bias for $\beta_{1}$ and $\beta_{3}$ and again poor CI coverage. JAV continued to be unbiased with correct CI coverage. SMC-FCS was somewhat biased, due to mis-specification of the models for $X_{1}|X_{2}$ and $X_{2}|X_{1}$, although CI coverage was only slightly below the nominal level for $\beta_{1}$ and $\beta_{3}$. When $X_{2}$ was normally distributed with mean a quadratic in $X_{1}$, FCS was again biased. JAV continued to be approximately unbiased. SMC-FCS was again somewhat biased, with CI coverage for $\beta_{3}$ approximately 88\%. With $X_{1}$ Bernoulli and $X_{2}|X_{1}$ normal, both JAV and SMC-FCS were unbiased, although SMC-FCS was slightly more efficient. Both had empirical CI coverge of approximately 95\% for both $\beta_{1}$ and $\beta_{3}$. It is important to note that here SMC-FCS is not equivalent to imputation a Bayesian model, and thus there is no guarantee that Rubin's rules will give asymptotically unbiased variance estimates. That the CI for $\beta_{3}$ from SMC-FCS had the correct coverage in this setting is thus encouraging. FCS was again biased. As expected, with $X_{2}$ log-normal given $X_{1}$ JAV continued to remain approximately unbiased while SMC-FCS had moderately large biases for $\beta_{1}$ and $\beta_{3}$, although CI coverage was only slightly below 95\%.

When $X_{1}$ and $X_{2}$ were MAR CC analysis was biased, due to the fact missingness was dependent on the outcome $Y$. FCS continued to be biased for all $X_{1},X_{2}$ distributions considered. With $X_{1},X_{2}$ bivariate normal, estimates from JAV had a small bias towards zero for $\beta_{3}$, but a larger bias for $\beta_{1}$. In contrast, SMC-FCS was unbiased and more efficient. In a number of simulated datasets with covariates log-normally distributed or with $X_{2}$ quadratic given $X_{1}$ the FCS algorithm created imputed datasets with extremely large imputed values of $X_{1}$ and $X_{2}$, resulting in a co-linearity error when attempting to fit the substantive model to the imputations. Consequently, for these scenarios results are shown for the subset of the simulated datasets for which estimates from all methods were obtained. JAV was approximately unbiased for $\beta_{3}$ when the covariates were log-normally distributed, but was substantially biased for $\beta_{1}$. SMC-FCS had a small bias for $\beta_{3}$ and a much smaller bias than JAV for $\beta_{1}$. With $X_{2}$ given $X_{1}$ normal with mean quadratic in $X_{1}$, both JAV and SMC-FCS were biased, but again biases for $\beta_{1}$ and $\beta_{3}$ were smaller for SMC-FCS. Lastly, with $X_{1}$ binary and $X_{2}$ either conditionally normal or log-normal, JAV had little bias for $\beta_{3}$, but had some bias for $\beta_{1}$. SMC-FCS was unbiased for both $\beta_{1}$ and $\beta_{3}$ when $X_{2}$ was conditionally normal given $X_{1}$. With $X_{2}$ log-normally distributed, SMC-FCS had a similar sized (but in the opposite direction) bias for $\beta_{1}$ as JAV, but was biased for $\beta_{3}$ whereas JAV was unbiased.

\begin{landscape}
\begin{table}
\caption{\label{linearRegInteractionResults}Simulation results - linear regression with interaction. Empirical mean (SD) of estimates of $\beta_{1}=1$ and $\beta_{3}=1$ from 1,000 simulations, using complete case analysis, standard FCS imputation (FCS), just another variable imputation (JAV), and SMC-FCS. Empirical coverage of nominal 95\% confidence intervals is also shown (Cov). Monte-Carlo errors for means and SDs are all less than 0.04 for $\beta_{1}$ and less than 0.02 for $\beta_{3}$. Monte-Carlo errors for confidence interval coverage are less than 1.6\%.}
\fbox{%
\begin{tabular}{*{10}{l}}
&  &  \multicolumn{2}{c}{Complete case} & \multicolumn{2}{c}{FCS} & \multicolumn{2}{c}{JAV} & \multicolumn{2}{c}{SMC-FCS} \\
$X_{1},X_{2}$ distribution & Parameter & Mean (SD) & Cov & Mean (SD) & Cov & Mean (SD) & Cov & Mean (SD) & Cov \\
\hline
MCAR \\
\hline
\multirow{2}{*}{$X_{1},X_{2}$ bivariate normal} & $\beta_{1}$ & 0.99 (0.55) & 94.5 & 1.46 (0.40) & 85.4 & 1.00 (0.53) & 95.1 & 0.98 (0.45) & 95.1 \\
& $\beta_{3}$ & 1.01 (0.23) & 94.9 & 0.76 (0.15) & 79.7 & 1.00 (0.23) & 95.4 & 1.01 (0.19) & 95.2 \\
\hline
\multirow{2}{*}{$X_{1},X_{2}$ bivariate log-normal} & $\beta_{1}$ & 0.99 (0.61) & 0.95 & 1.70 (0.58) & 71.3 & 1.01 (0.60) & 94.5 & 0.81 (0.56) & 93.0 \\
& $\beta_{3}$ & 1.01 (0.22) & 96.1 & 0.69 (0.24) & 63.2 & 1.00 (0.22) & 95.5 & 1.08 (0.22) & 92.2 \\
\hline
\multirow{2}{*}{$X_{1}$ normal, $X_{2}|X_{1}\sim N((X_{1}-2)^{2},2)$} & $\beta_{1}$ & 1.01 (0.51) & 95.6 & 2.06 (0.59) & 41.5 & 1.03 (0.50) & 94.8 & 1.09 (0.53) & 92.3 \\
& $\beta_{3}$ & 1.00 (0.13) & 94.8 & 0.75 (0.23) & 63.1 & 1.00 (0.13) & 94.1 & 1.09 (0.14) & 87.6 \\
\hline
\multirow{2}{*}{$X_{1}$ Bernoulli, $X_{2}|X_{1}$ normal} & $\beta_{1}$ & 0.99 (0.24) & 94.4 & 1.11 (0.21) & 91.4 & 1.00 (0.23) & 94.2 & 0.99 (0.22) & 94.1 \\
& $\beta_{3}$ & 0.99 (0.20) & 94.2 & 0.82 (0.15) & 84.2 & 0.98 (0.20) & 94.9 & 0.99 (0.17) & 94.0 \\
\hline
\multirow{2}{*}{$X_{1}$ Bernoulli, $X_{2}|X_{1}$ log-normal} & $\beta_{1}$ & 1.02 (0.75) & 95.0 & 1.72 (0.63) & 79.0 & 1.05 (0.74) & 95.1 & 1.28 (0.66) & 92.7 \\
& $\beta_{3}$ & 0.99 (0.28) & 94.1 & 0.72 (0.24) & 79.0 & 0.98 (0.28) & 94.1 & 0.89 (0.25) & 91.3 \\
\hline
MAR \\
\hline
\multirow{2}{*}{$X_{1},X_{2}$ bivariate normal} & $\beta_{1}$ & 0.96 (0.50) & 94.1 & 1.61 (0.37) & 82.2 & 1.36 (0.60) & 87.9 & 1.02 (0.45) & 95.5 \\
& $\beta_{3}$ & 0.79 (0.24) & 84.7 & 0.64 (0.12) & 57.5 & 0.93 (0.30) & 94.0 & 0.99 (0.19) & 96.1 \\
\hline
\multirow{2}{*}{$X_{1},X_{2}$ bivariate log-normal*} & $\beta_{1}$ & 0.99 (0.84) & 94.8 & 2.49 (1.01) & 42.8 & 1.70 (1.14) & 88.5 & 0.93 (0.97) & 94.1 \\
& $\beta_{3}$ & 0.77 (0.40) & 91.0 & 0.19 (0.39) & 29.9 & 1.01 (0.55) & 94.1 & 1.06 (0.48) & 90.0 \\
\hline
\multirow{2}{*}{$X_{1}$ normal, $X_{2}|X_{1}\sim N((X_{1}-2)^{2},2)$**} & $\beta_{1}$ & 0.83 (0.46) & 94.8 & 2.36 (1.36) & 33.7 & 1.68 (0.63) & 82.4 & 1.27 (0.54) & 94.4 \\
& $\beta_{3}$ & 0.85 (0.15) & 83.6 & 0.15 (0.30) & 22.6 & 1.16 (0.20) & 87.4 & 1.10 (0.20) & 89.9 \\
\hline
\multirow{2}{*}{$X_{1}$ Bernoulli, $X_{2}|X_{1}$ normal} & $\beta_{1}$ & 0.86 (0.21) & 90.0 & 1.11 (0.21) & 93.3 & 1.15 (0.22) & 88.4 & 1.00 (0.22) & 94.5 \\
& $\beta_{3}$ & 0.81 (0.19) & 84.6 & 0.79 (0.14) & 83.4 & 0.97 (0.22) & 93.7 & 0.98 (0.17) & 95.1 \\
\hline
\multirow{2}{*}{$X_{1}$ Bernoulli, $X_{2}|X_{1}$ log-normal} & $\beta_{1}$ & 1.04 (0.79) & 95.3 & 1.79 (0.74) & 84.7 & 0.83 (0.95) & 95.7 & 1.21 (0.76) & 93.8 \\
& $\beta_{3}$ & 0.78 (0.30) & 90.7 & 0.71 (0.27) & 84.4 & 1.00 (0.37) & 92.8 & 0.92 (0.28) & 93.5 \\
\multicolumn{2}{c}{* results based on 994 simulations} \\
\multicolumn{2}{c}{** results based on 959 simulations}
\end{tabular}}
\end{table}
\end{landscape}

\subsection{Cox proportional hazards models}
\label{CoxSimulations}
Lastly, we performed simulations for imputing missing covariates with a Cox proportional hazards model. Recently \cite{White2009} derived approximate results to inform the choice of imputation model in this context. They recommended including the event indicator and the Nelson-Aalen estimate of the marginal cumulative hazard function as covariates in imputation models. Their simulation results showed that this approach generally worked well for imputing normally distributed covariates, except when the covariate effects were large, when some attenuation towards the null occurred.

\subsubsection{Simulation setup}
Survival times were simulated with hazard function $h(t|X)=0.002\exp(\beta_{1} X_{1}+\beta_{2} X_{2})$ with $\beta_{1}=\beta_{2}=1$. Censoring times were generated from an exponential distribution with hazard 0.002. We simulated $X_{1}$ from a Bernoulli distribution with probability 0.5, and $X_{2}|X_{1} \sim N(X_{1},1)$. Values in $X_{1}$ and $X_{2}$ were made (independently) missing completely at random, with probability of observation 0.7. We performed simulations with $n=1,000$ subjects and also with $n=100$ subjects.

\subsubsection{Estimation methods}
For each simulated dataset we first estimated $\beta$ by fitting the Cox proportional hazards model to the complete cases. Next we multiply imputed the missing values in $X_{1}$ and $X_{2}$ using FCS (10 imputations). A linear regression imputation model was used for $X_{2}$ and a logistic regression model for $X_{1}$. Following the recommendations of \cite{White2009}, we included the event indicator $D$ and the Nelson-Aalen estimate of the marginal cumulative hazard as covariates in both imputation models (FCS). Lastly we estimated $\beta$ using SMC-FCS as described in Section \ref{proportionalHazardsModels}, assuming a logistic regression model for $X_{1}|X_{2}$ and a linear regression model for $X_{2}|X_{1}$. As described previously, the SMC-FCS algorithm involves taking draws from the posterior distribution of the parameter $\psi$ in the substantive model. For Cox's proportional hazards model $\psi=(\beta,H_{0}(.))$, where $H_{0}(.)$ is an infinite dimensional parameter representing the arbitrary baseline hazard function. It is unclear how a draw can be made from the posterior distribution of $H_{0}(.)$, and indeed whether Rubin's rules can be expected to give asymptotically unbiased variance estimates in a semi-parametric model. In our simulation study we allowed for uncertainty in $\beta$ by drawing a new value from a (bivariate) normal distribution with mean equal to the current estimate of $\beta$ and with covariance matrix based on the usual `observed' data information matrix. We then updated $H_{0}(.)$ using the usual Breslow estimator, conditioning on the newly drawn value of $\beta$.

\subsubsection{Results}
Table \ref{coxResultsTable} shows the results from the 1,000 simulations. CC is consistent here, since missingness is completely at random. However, with $n=100$ CC showed some upward finite sample bias for both $\beta_{1}$ and $\beta_{2}$. In accordance with the results of White and Royston, FCS resulted in somewhat biased estimates, with the bias larger for the coefficient corresponding to the continuous covariate, although confidence interval coverage for both $\beta_{1}$ and $\beta_{2}$ was approximately 95\%. SMC-FCS, like CC, showed some slight upward bias, but was somewhat more efficient. Of interest was that the confidence intervals had correct coverage, despite the fact that our implementation ignores uncertainty in the baseline hazard function.

For $n=1,000$, CC was essentially unbiased. The biases of FCS were larger than for $n=100$, which is due to the fact that the finite sample bias, which acted in the opposite direction to the bias caused by the approximation used in the FCS approach, had largely disappeared. Consequently, confidence interval coverage was below the nominal 95\% level, with coverage for $\beta_{2}$ particularly poor at 47\%. In contrast, SMC-FCS was unbiased and had correct confidence interval coverage.

\begin{table}
\caption{\label{coxResultsTable}Cox proportional hazards outcome model simulation results. Empirical mean (SD) of estimates of $\beta_{1}=1$ and $\beta_{2}=1$ from 1,000 simulations, using complete case analysis, multiple imputation of $X_{1}$ and $X_{2}$ using FCS with the event indicator and Nelson-Aalen marginal baseline cumulative hazard function as covariates (FCS), and SMC-FCS. Empirical coverage of nominal 95\% confidence intervals is also shown (Cov). Monte-Carlo errors in means and SDs are no more than 0.02 for $n=100$ and 0.005 for $n=1000$.}
\centering
\fbox{%
\begin{tabular}{lllllll}
Parameter & \multicolumn{2}{c}{Complete case} & \multicolumn{2}{c}{FCS} & \multicolumn{2}{c}{SMC-FCS} \\
& Mean (SD) & Cov & Mean (SD) & Cov & Mean (SD) & Cov  \\
\hline
$n=100$ \\
$\beta_{1}=1$  & 1.04 (0.47) & 95.6 & 0.97 (0.36) & 96.7 & 1.02 (0.41) & 94.9  \\
$\beta_{2}=1$  & 1.05 (0.26) & 95.6 & 0.88 (0.17) & 94.6 & 1.05 (0.21) & 94.5  \\
\hline
$n=1,000$ \\
$\beta_{1}=1$  & 1.000 (0.129) & 95.2 & 0.896 (0.105) & 89.7 & 1.000 (0.116) & 94.9 \\
$\beta_{2}=1$  & 1.007 (0.070) & 94.8 & 0.865 (0.050) & 47.2 & 1.006 (0.058) & 95.1 \\
\end{tabular}}
\end{table}

\section{Analysis of data from ADNI}
\label{adni}
Table \ref{adniResults} shows the estimated log hazard ratios from the substantive model fitted to the $n=127$ complete cases (of whom 61 converted to AD). This showed borderline evidence of an association between CSF A$\beta_{1-42}$ and hazard of conversion, and borderline significant evidence of curvature in the association, in agreement with the findings of \cite{Jack2010}. The estimated association suggests that increasing A$\beta_{1-42}$ is associated with increased hazard of conversion up until a value of $\approx 14$ ng/mL, after which hazard decreases. There was evidence that increased levels of P-tau were associated with increased hazard of conversion. Contrary to what we expected, having a mother or father with AD was suggestive of lower hazard of conversion to AD, although neither coefficient was statistically significant. Hippocampal volume was the strongest predictor of hazard (measured by statistical significance), with larger volumes associated with lower hazard of conversion. This is consistent with the findings of previous studies which have found that the hippocampus is one of the earliest structures of the brain to undergo atrophy during AD.

Next we used FCS MI to impute the partially observed baseline variables. 50 imputations were used. Continuous variables were imputed using linear regression models while binary variables were imputed using logistic regressions. To incorporate the censored time to conversion outcome we followed the recommendations of \cite{White2009} and included the event indicator and marginal Nelson-Aalen cumulative hazard function as covariates in the imputation models. We passively imputed the A$\beta_{1-42}^{2}$ term in the imputed datasets. The FCS estimate of A$\beta_{1-42}^{2}$ is smaller in magnitude than the CC estimate (Table \ref{adniResults}). This is consistent with the simulation results of Section \ref{quadraticSimulations}, which showed that linear imputation of variables for substantive models which include quadratic effects of the variable leads to attenuation in the estimate of curvature. The coefficient for the linear A$\beta_{1-42}$ term is also much smaller and no longer statistically significant. The estimated coefficient for P-tau is much smaller. The negative association between hippocampal volume and hazard of conversion remained. The coefficients for family history of AD changed by a proportionately large amount. Further investigation revealed that those with family history of AD were much more likely to have CSF variables measured, and that the dependence of hazard of conversion to AD on family history of AD varied strongly (in a model without the CSF variables) according to whether or not the CSF variables were measured. This means that the assumption required for validity of complete case analysis failed for the reduced model without the CSF variables, and this is the likely cause of the large change in the coefficients for family history of AD. Standard errors were considerably smaller, consistent with the gain in information through inclusion of subjects with some missing values into the analysis.

Lastly, we imputed using SMC-FCS, again using 50 imputations. Here we assumed linear regression covariate models for partially observed continuous variables and logistic regressions for partially observed binary variables. Comparing the estimates from SMC-FCS with complete case and passive FCS, we see that the linear and quadratic coefficients of A$\beta_{1-42}$ are much closer to the complete case estimates, with the statistical significance of the quadratic coefficient preserved  (Table \ref{adniResults}). The estimated coefficient for P-tau is similar to that obtained in the FCS analysis. For the other coefficients the SMC-FCS estimates and CIs are similar to those from FCS. In conclusion, consistent with our earlier simulation results, the results of this analysis suggest that ignoring the quadratic association at the imputation stage leads to attenuation in the corresponding coefficient. In contrast, imputation of the partially observed covariates using SMC-FCS preserved a quadratic association between CSF A$\beta_{1-42}$ seen in the complete cases. Furthermore, use of MI has here lead to practically important improvements in the precision of estimated associations, compared to CC.

\begin{table}
\caption{\label{adniResults}Estimates of log hazard ratios (standard errors) for Cox proportional hazards model relating hazard of conversion to AD to baseline risk factors. Estimates based on complete case, FCS imputation, and SMC-FCS.}
\centering
\fbox{%
\begin{tabular}{llll}
& (n=127) & \multicolumn{2}{c}{(n=382)} \\
Variable & Complete case & FCS & SMC-FCS \\
\hline
A$\beta_{1-42}$ (ng/mL) & 0.31 (0.19) & 0.08 (0.10) & 0.28 (0.16) \\
A$\beta_{1-42}^{2}$ (ng$^{2}$/mL$^{2}$) & -0.011 (0.005) & -0.004 (0.003) & -0.010 (0.005) \\
log(Tau) (log pg/mL) & -0.60 (0.47) & -0.23 (0.37) & -0.17 (0.38) \\
log(P-tau) (log pg/mL) & 1.29 (0.51) & 0.52 (0.38) & 0.47 (0.23) \\
Mother had AD & -0.61 (0.32) & -0.15 (0.22) & -0.14 (0.21) \\
Father had AD & -1.07 (0.68) & -0.22 (0.35) & -0.26 (0.38) \\
Intracranial volume (cm$^3$) & 0.0005 (0.0010) & 0.0010 (0.0007) & 0.0011 (0.0007)\\
Hippocampal volume (cm$^3$) & -0.64 (0.17) & -0.47 (0.10) & -0.50 (0.10) \\
APOE4 positive & -0.06 (0.30) & 0.31 (0.22) & 0.32 (0.22) \\
\end{tabular}}
\end{table}

\section{Multiple imputation of covariates in practice}
\label{exploratoryAnalyses}
Our developments thus far have assumed that at the imputation stage we have a single correctly specified substantive model $f(Y|X,Z,\psi)$. Often we may not know in advance of analysing the data what is an appropriate model for the outcome $Y$ of interest given the covariates. One of the apparent advantages of using MI is that once a set of imputed datasets have been generated, a number of different substantive models can be fitted and compared. Of course the validity of estimates from different models fitted to a set of multiple imputations depends on whether the imputation model is correctly specified. In practice all imputation models are likely to be mis-specified to some extent, but biases may be small provided imputation models preserve those features of the data which are subsequently investigated. It is therefore unrealistic in practice to expect a single set of multiple imputations to be suitable for all possible subsequent types of analysis.

When a number of putative substantive models for the outcome $Y$ are of interest, the SMC-FCS algorithm could be used to impute the partially observed covariates assuming a general model for $f(Y|X,Z,\psi)$ which contains as special cases the various putative substantive models. This approach would mean the covariate models used would be compatible with this larger model, and semi-compatible with those substantive models nested within it. This advice follows that given by others (e.g. \cite{Meng:1994,Schafer:1997}) for application of MI in general, whereby imputation models are used which are rich and do not impose assumptions which are subsequently to be relaxed in substantive models. For example, if based on contextual knowledge and preliminary data analysis it is thought that two partially covariates may interact in their effect on $Y$, one could impute under a model $f(Y|X,Z,\psi)$ which includes the corresponding interaction.

As noted in Section \ref{introduction} in many settings auxiliary variables $V$ may be available, which although not involved in the substantive model, may be useful for inclusion in imputation models in order to improve efficiency (by virtue of their association with variables being imputed) or to increase the plausibility of the MAR assumption. The notion of compatibility between imputation and substantive models does not then apply, since the two models involve different sets of variables. However,fully observed auxiliary variables $Z$ could be included as additional fully observed covariates (i.e. incorporated as part of $Z$).

\section{Discussion}
\label{discussion}
Multiple imputation is an attractive approach for handling missingness in covariates of regression models. In many settings standard choices of covariate imputation models are compatible with the substantive model, rendering our proposed approach unnecessary. However for certain substantive models default imputation models in MI software may be incompatible and therefore mis-specified (assuming the substantive model is correctly specified). This is particularly likely to be the case for substantive models which contain non-linear covariate effects or interactions. Our proposed modification of the popular FCS approach to MI ensures that each covariate is imputed from a model which is compatible with the substantive model. Although compatibility does not guarantee the imputation models are correctly specified, it ensures that the imputation and substantive models do not make conflicting assumptions which may induce bias in parameter estimates.

In special cases SMC-FCS is equivalent to imputation from a Bayesian joint model, and thus inherits the latter's properties. More generally, if the covariate models used in SMC-FCS are mutually compatible and correctly specified we conjecture that the resulting estimator is consistent, which is supported by our simulation results. Further, in these cases confidence interval coverage for estimates from SMC-FCS attained nominal coverage, despite the lack of equivalence to imputation from a Bayesian joint model. In simulations in which the covariate models were mis-specified, estimates from SMC-FCS were still less biased than those from what might be considered `standard FCS'.

For linear substantive models which contain non-linear covariate effects or interactions, the `just another variable' (JAV) approach is attractive, and is consistent if data are missing completely at random. This holds irrespective of the joint distribution of the outcome and covariates. However when data are MAR, or for other substantive model types such as logistic regression, we and Seaman \etal \cite{Seaman2012} have shown that JAV gives biased estimates. At least in our limited simulation study, the polynomial combination method recently proposed by van Buuren \cite{vanBuuren2012} was superior to JAV, with less bias and coverage closer to the nominal level. A limitation of this approach however is that it is only applies to imputation of covariates which have a quadratic association with outcome, and it is unclear whether it can be generalised to substantive models other than linear regression.

Relative to standard FCS MI, SMC-FCS is more computationally intensive because of the use of rejection sampling to sample from the required densities. For example, the SMC-FCS algorithm took six times longer than standard FCS to create 10 imputations for a simulated dataset from the first simulation scenario (linear regression with quadratic covariate effects). The acceptance rate of the rejection sampler will be low when the target density $f(X_{j}|X_{-j},Z,Y)$ differs substantially from the candidate density $f(X_{j}|X_{-j},Z)$. This will occur if a subject has an outcome value $Y$ which is unlikely to have occurred given the values of $X_{-j}$ and $Z$. However our experience thus far in simulation studies has been that this has not been an issue. Furthermore, as for standard FCS MI, additional work is needed to understand the statistical properties of the SMC-FCS algorithm. In some settings substantive models may be fitted to imputed datasets for a number of different outcomes, and a limitation of our approach is that imputation models are defined with respect to a single (possibly multivariate) outcome variable.

We note that compatibility between imputation and substantive models is closely related to the concept of congeniality defined by Meng \cite{Meng:1994}. We chose not to adopt this term because Meng's definition of congeniality depends additionally on specification of incomplete and complete data `procedures' which give asymptotically equivalent inferences to those under a Bayesian model. Further, in many cases (e.g. when a logistic regression model is used to impute a covariate) SMC-FCS is not equivalent to imputation from a joint model, and so would not satisfy Meng's definition of congeniality. Lastly, the setup adopted by Meng assumed that covariates are fully observed.

In this paper we have assumed that the outcome is fully observed. In the absence of auxiliary variables subjects with missing outcome provide little or no additional information regarding the substantive model parameters \cite{Little:1992}, such that imputation of missing outcomes may not be beneficial. Nevertheless, the SMC-FCS algorithm can be readily extended to impute missing outcome values by imputing from the assumed substantive model.

A Stata program implementing SMC-FCS for linear, logistic, and Cox proportional hazards models of interest is available for free download from www.missingdata.org.uk.

\section{Acknowledgments}
J. Bartlett was supported by a grant from the ESRC Follow-On Funding scheme (RES-189-25-0103) and MRC grant G0900724. S. Seaman and I. White were supported through a Medical Research Council grant (MC\_US\_A030\_0015) and unit programme (U105260558). J. Carpenter was supported by ESRC Research Fellowship RES-063-27-0257.

Data used in the preparation of this article were obtained from the Alzheimer's Disease Neuroimaging
Initiative (ADNI) database (adni.loni.ucla.edu). The ADNI was launched in 2003 by the National Institute on
Aging (NIA), the National Institute of Biomedical Imaging and Bioengineering (NIBIB), the Food and Drug
Administration (FDA), private pharmaceutical companies and non-profit organizations, as a \$60 million, 5-
year public-private partnership. The primary goal of ADNI has been to test whether serial magnetic
resonance imaging (MRI), positron emission tomography (PET), other biological markers, and clinical and
neuropsychological assessment can be combined to measure the progression of mild cognitive impairment
(MCI) and early Alzheimer's disease (AD). Determination of sensitive and specific markers of very early AD
progression is intended to aid researchers and clinicians to develop new treatments and monitor their
effectiveness, as well as lessen the time and cost of clinical trials.

The Principal Investigator of this initiative is Michael W. Weiner, MD, VA Medical Center and University of
California – San Francisco. ADNI is the result of efforts of many co-investigators from a broad range of
academic institutions and private corporations, and subjects have been recruited from over 50 sites across
the U.S. and Canada. The initial goal of ADNI was to recruit 800 adults, ages 55 to 90, to participate in the
research, approximately 200 cognitively normal older individuals to be followed for 3 years, 400 people with
MCI to be followed for 3 years and 200 people with early AD to be followed for 2 years." For up-to-date
information, see www.adni-info.org.

Data collection and sharing for this project was funded by the Alzheimer's Disease Neuroimaging Initiative
(ADNI) (National Institutes of Health Grant U01 AG024904). ADNI is funded by the National Institute on
Aging, the National Institute of Biomedical Imaging and Bioengineering, and through generous contributions
from the following: Abbott; Alzheimer's Association; Alzheimer's Drug Discovery Foundation; Amorfix Life
Sciences Ltd.; AstraZeneca; Bayer HealthCare; BioClinica, Inc.; Biogen Idec Inc.; Bristol-Myers Squibb
Company; Eisai Inc.; Elan Pharmaceuticals Inc.; Eli Lilly and Company; F. Hoffmann-La Roche Ltd and its
affiliated company Genentech, Inc.; GE Healthcare; Innogenetics, N.V.; IXICO Ltd.; Janssen Alzheimer
Immunotherapy Research \& Development, LLC.; Johnson \& Johnson Pharmaceutical Research \&
Development LLC.; Medpace, Inc.; Merck \& Co., Inc.; Meso Scale Diagnostics, LLC.; Novartis
Pharmaceuticals Corporation; Pfizer Inc.; Servier; Synarc Inc.; and Takeda Pharmaceutical Company. The
Canadian Institutes of Health Research is providing funds to support ADNI clinical sites in Canada. Private
sector contributions are facilitated by the Foundation for the National Institutes of Health (www.fnih.org).
The grantee organization is the Northern California Institute for Research and Education, and the study is coordinated by the Alzheimer's Disease Cooperative Study at the University of California, San Diego. ADNI
data are disseminated by the Laboratory for Neuro Imaging at the University of California, Los Angeles. This
research was also supported by NIH grants P30 AG010129 and K01 AG030514.


\begin{thebibliography}{10}

\bibitem{Kenward/Carpenter:2007}
M~G Kenward and J~R Carpenter.
\newblock Multiple imputation: current perspectives.
\newblock {\em Statistical Methods in Medical Research}, 16:199--218, 2007.

\bibitem{White2010a}
I.~R. White and J.~B. Carlin.
\newblock {Bias and efficiency of multiple imputation compared with
  complete-case analysis for missing covariate values}.
\newblock {\em Statistics in Medicine}, 28:2920--2931, 2010.

\bibitem{Rubin:1987}
D~B Rubin.
\newblock {\em {Multiple imputation for nonresponse in surveys}}.
\newblock New York: Wiley, 1987.

\bibitem{White2011}
I~R White, P~Royston, and A~M Wood.
\newblock Multiple imputation using chained equations: issues and guidance for
  practice.
\newblock {\em Statistics in Medicine}, 30:377--399, 2011.

\bibitem{Buuren:2007}
S~van Buuren.
\newblock Multiple imputation of discrete and continuous data by fully
  conditional specification.
\newblock {\em Statistical Methods in Medical Research}, 16:219--242, 2007.

\bibitem{Meng:1994}
X~L Meng.
\newblock Multiple-imputation inferences with uncongenial sources of input
  (with discussion).
\newblock {\em Statistical Science}, 10:538--573, 1994.

\bibitem{Schafer:1997}
J~L Schafer.
\newblock {\em {Analysis of incomplete multivariate data}}.
\newblock London: Chapman and Hall, 1997.

\bibitem{Collins2001}
L.~M. Collins, J.~L. Schafer, and C.~Kam.
\newblock A comparison of inclusive and restrictive strategies in modern
  missing data procedures.
\newblock {\em {Psychological Methods}}, 6:330--351, 2001.

\bibitem{Schafer2003}
J.~L. Schafer.
\newblock {Multiple imputation in multivariate problems when the imputation and
  analysis models differ}.
\newblock {\em Statistica Neerlandica}, 57:19--35, 2003.

\bibitem{Seaman2012}
S.~R. Seaman, J.~W. Bartlett, and I.~R. White.
\newblock {Multiple imputation of missing covariates with non-linear effects
  and interactions: an evaluation of statistical methods}.
\newblock {\em BMC Medical Research Methodology}, 12:46, 2012.

\bibitem{Jack2010}
C~R {Jack Jr}, H~J Wiste, P~Vemuri, S~D Weigand, M~L Senjem, G~Zeng, M~A
  Bernstein, J~L Gunter, V~S Pankratz, P~S Aisen, M~W Weiner, R~C Petersen, L~M
  Shaw, J~Q Trojanoswski, D~S Knopman, and {Alzheimer's Disease Neuroimaging
  Initiative}.
\newblock {Brain beta-amyloid measures and magnetic resonance imaging atrophy
  both predict time-to-progression from mild cognitive impairment to
  Alzheimer's disease}.
\newblock {\em Brain}, 133:3336--3348, 2010.

\bibitem{Rubin:1976}
D~B Rubin.
\newblock {Inference and missing data}.
\newblock {\em Biometrika}, 63:581--592, 1976.

\bibitem{Liu2012}
J~Liu, A~Gelman, J~Hill, and Y~Su.
\newblock On the stationary distribution of iterative imputations.
\newblock {\em arXiv}, page 1012.2902v2, 2012.

\bibitem{Arnold1989}
B~C Arnold and S~J Press.
\newblock Compatible conditional distributions.
\newblock {\em Journal of the American Statistical Association}, 84:152--156,
  1989.

\bibitem{Hippel2009}
P~T von Hippel.
\newblock How to impute interactions, squares, and other transformed variables.
\newblock {\em Sociological Methodology}, 39:265--291, 2009.

\bibitem{White2009}
I.~R. White and P.~Royston.
\newblock {Imputing missing covariate values for the Cox model}.
\newblock {\em Statistics in Medicine}, 28:1982--1998, 2009.

\bibitem{Ibrahim:1999}
J~G Ibrahim, M~H Chen, and S~R Lipsitz.
\newblock {Monte-Carlo EM for missing covariates in parametric regression
  models}.
\newblock {\em Biometrics}, 55:591--596, 1999.

\bibitem{vanBuuren2006}
S~{van Buuren}, J~P~L Brand, C~G~M {Groothuis-Oudshoorn}, and D~B Rubin.
\newblock Fully conditional specification in multivariate imputation.
\newblock {\em Journal of Statistical Computation and Simulation},
  76:1049--1064, 2006.

\bibitem{Gelman2004}
A.~Gelman, J.~B. Carlin, H.~S. Stern, and D.~B. Rubin.
\newblock {\em {Bayesian Data Analysis}}.
\newblock Boca Raton: Chapman \& Hall/CRC, 2nd edition, 2004.

\bibitem{Rathouz2007}
P~J Rathouz.
\newblock Identifiability assumptions for missing covariate data in failure
  time regression models.
\newblock {\em Biostatistics}, 8:345--356, 2007.

\bibitem{vanBuuren2012}
S~van Buuren.
\newblock {\em {Flexible imputation of missing data}}.
\newblock Boca Raton: Chapman \& Hall/CRC, 2012.

\bibitem{Little:1992}
R~J~A Little.
\newblock {Regression With Missing X's: A Review}.
\newblock {\em Journal of the American Statistical Association}, 87:1227--1237,
  1992.

\end{thebibliography}
\end{document}